\documentclass[sigconf]{acmart}
\AtBeginDocument{%
  }

\copyrightyear{2026}
\acmYear{2026}
\setcopyright{cc}
\setcctype{by}
\acmConference[SIGIR '26]{Proceedings of the 49th International ACM SIGIR Conference on Research and Development in Information Retrieval}{July 20--24, 2026}{Melbourne, VIC, Australia}
\acmBooktitle{Proceedings of the 49th International ACM SIGIR Conference on Research and Development in Information Retrieval (SIGIR '26), July 20--24, 2026, Melbourne, VIC, Australia}
\acmDOI{10.1145/3805712.3808401}
\acmISBN{979-8-4007-2599-9/2026/07}

\usepackage{enumitem}
\usepackage{multirow}
\usepackage{amsmath}
\usepackage{amsthm}
\usepackage{subfigure}
\usepackage{xcolor, colortbl}
\usepackage{graphicx}
\usepackage{subcaption}
\usepackage{caption}

\definecolor{LightBlue}{rgb}{0.88,1,1}
\definecolor{LightGreen}{rgb}{0.88,1,0.88}
\definecolor{LightBrown}{rgb}{0.99,0.95,0.81}
\definecolor{CSOrange}{HTML}{f39c12}
\definecolor{CSGreen}{HTML}{52be80}
\definecolor{CSBlue}{HTML}{5dade2}

\begin{document}



\title[CS3: Capability Synergy for Two-Tower Recommendation]{CS3: Efficient Online Capability Synergy for \\ Two-Tower Recommendation}


\author{Lixiang Wang}
\authornote{Authors contributed equally to this research.}
\affiliation{
  \institution{Kuaishou Technology}
  \city{Hangzhou}
  \country{China}
}
\email{wanglixiang03@kuaishou.com}

\author{Shaoyun Shi}
\authornotemark[1]
\affiliation{
  \institution{Kuaishou Technology}
  \city{Hangzhou}
  \country{China}
}
\email{shisy13@outlook.com}

\author{Peng Wang}
\authornotemark[1]
\affiliation{
  \institution{Kuaishou Technology}
  \city{Hangzhou}
  \country{China}
}
\email{wangpeng16@kuaishou.com}

\author{Wenjin Wu}
\authornote{Corresponding author.}
\affiliation{
  \institution{Kuaishou Technology}
  \city{Beijing}
  \country{China}
}
\email{wuwenjin@kuaishou.com}

\author{Peng Jiang}
\affiliation{
  \institution{Kuaishou Technology}
  \city{Beijing}
  \country{China}
}
\email{jiangpeng@kuaishou.com}

\begin{abstract}


To balance effectiveness and efficiency in recommender systems, multi-stage pipelines commonly use lightweight two-tower models for large-scale candidate retrieval. However, the isolated two-tower architecture restricts representation capacity, embedding-space alignment, and cross-feature interactions. Existing solutions such as late interaction and knowledge distillation can mitigate these issues, but often increase latency or are difficult to deploy in online learning settings. We propose \textit{Capability Synergy} (\textbf{CS3}), an efficient online framework that strengthens two-tower retrievers while preserving real-time constraints. CS3 introduces three mechanisms: (1) \textit{Cycle-Adaptive Structure} for self-revision via adaptive feature denoising within each tower; (2) \textit{Cross-Tower Synchronization} to improve alignment through lightweight mutual awareness between towers; and (3) \textit{Cascade-Model Sharing} to enhance cross-stage consistency by reusing knowledge from downstream models. CS3 is plug-and-play with diverse two-tower backbones and compatible with online learning. Experiments on three public datasets show consistent gains over strong baselines, and deployment in a large-scale advertising system yields up to 8.36\% revenue improvement across three scenarios while maintaining ms-level latency.

\end{abstract}

\ccsdesc[500]{Information systems~Recommender systems}

\keywords{Two-Tower, Recommendation, Capability Synergy, Online Learning}



\maketitle


\section{Introduction}

To balance effectiveness and efficiency, industrial recommender systems are typically implemented as multi-stage pipelines, where early-stage retrieval must operate under strict latency budgets\cite{LiuXOS17,CovingtonAS16YouTubeDNN,GrbovicC18}.
The two-tower model is a widely used architecture for large-scale candidate retrieval\cite{dssm2013,CovingtonAS16YouTubeDNN,senet}. It encodes users and items with two separate networks and computes relevance via dot product or cosine similarity. This design enables efficient serving: user embeddings can be computed once and reused, item embeddings can be precomputed and cached, and approximate nearest-neighbor search can be accelerated by libraries such as FAISS\cite{douze2025faisslibrary}.

Despite these advantages, the isolated two-tower structure limits retrieval quality in practice. First, \textbf{capacity} is constrained by the lightweight, modular design of each tower\cite{wang2020COLD,ma2021FSCD}. Second, \textbf{alignment} between user and item embedding spaces is difficult to maintain because the towers do not interact before similarity computation; relying solely on the training objective for alignment is often insufficient at industrial scale and becomes more challenging under online learning, where new items and feedback continuously shift the data distribution\cite{WangYM000M22DirectAU}. Third, \textbf{cross-stage consistency} is weak: retrievers cannot model rich user--item cross features\cite{WangFFW17DeepCross,BianWRPZXSZCMLX22CAN,YuYWZMLJK22CFFNN} or target-aware sequential signals\cite{SIM,TWIN,SDIM,ETA} that are routinely exploited by downstream rankers, creating a structural gap that hinders end-to-end optimization.

Prior work typically improves two-tower models along a single axis, e.g., introducing implicit cross-tower interactions\cite{dat,IntTower} or distilling knowledge from stronger teachers\cite{TangW18, reddi2021rankdistil}. However, these approaches often add overhead and, more importantly, do not provide a \emph{unified} mechanism to jointly address (i) within-tower refinement, (ii) explicit cross-tower coordination, and (iii) ranker-to-retriever knowledge reuse—an omission that is particularly limiting in online learning systems\cite{DBLP:conf/recsys/0004ZZWZTZZWWC22,cao2024momentcrossnextgenerationrealtimecrossdomain}, where updates must be frequent yet the serving budget remains tight.

\begin{figure*}[t]
\vspace{-1em}
\centering
\includegraphics[width=0.9\linewidth]{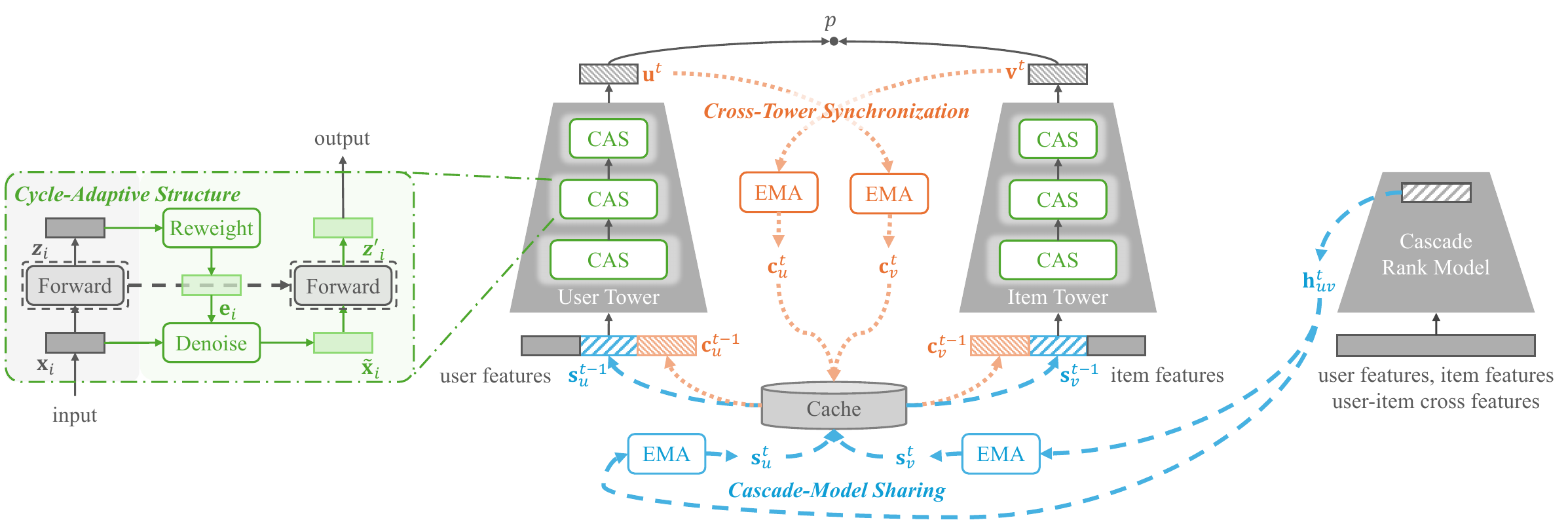}
\vspace{-0.5em}
\caption{An overview of the proposed CS3 framework. 
} 
\label{fig:CS3} 
\vspace{-1em}
\end{figure*}

In this paper, we propose \textit{Capability Synergy} (\textbf{CS3}), a deployable paradigm for strengthening two-tower retrievers in online, multi-stage recommender systems. CS3 frames retrieval enhancement as \emph{capability exchange} across three levels: self-revision within each tower, synchronization between towers, and knowledge sharing from downstream cascade models. This framing aligns with our abstract: CS3 is designed to be plug-and-play with diverse two-tower backbones while preserving millisecond-level latency and online learning compatibility. We summarize our contributions as:
\begin{itemize}[leftmargin=*]
\item \textbf{Framework Innovation}. We introduce CS3, a unified capability-synergy paradigm that connects within-tower refinement, cross-tower alignment, and cross-stage knowledge reuse, avoiding “single-trick” enhancements that are prone to incremental gains.
\item \textbf{Synergistic Modules}. We instantiate CS3 with three lightweight modules -- Cycle-Adaptive Structure (CAS), Cross-Tower Synchronization (CTS), and Cascade-Model Sharing (CMS) -- that can be integrated into standard two-tower architectures with minimal system changes.
\item \textbf{Industrial-Scale Validation}. We implement CS3 under strict latency constraints in an online learning advertising system, and demonstrate consistent improvements via offline experiments on three public datasets and online A/B tests.
\end{itemize}

\vspace{-0.5em}
\section{Methodology}

Figure~\ref{fig:CS3} illustrates the proposed CS3 framework. CS3 is compatible with various two-tower architectures. For exposition, we use a standard two-tower model where the preference score between a user $u \in \mathcal{U}$ and an item $v \in \mathcal{V}$ is computed as the dot product of their embeddings, i.e., $p=\mathbf{u}^\top \mathbf{v}$. We then describe the three core components of CS3 and its online implementation.

\vspace{-0.5em}
\subsection{Cycle-Adaptive Structure}

A single tower in a two-tower model learns user/item representations. Motivated by cyclic refinement and denoising in RecycleNet~\cite{RecycleNet} and diffusion models~\cite{DDPM,DDIM}, we propose \textit{Cycle-Adaptive Structure} (CAS) to replace the original fully connected (FC) layer in each tower (green part in Figure~\ref{fig:CS3}). CAS performs lightweight self-correction via one refinement cycle composed of \textit{pre-forward}, \textit{adaptive reweighting}, and \textit{cycle-forward}.

\textbf{Pre-Forward.} Let $\mathbf{x}_i\in\mathbb{R}^{d_i}$ be the input of layer $i$. Then
\begin{equation}
    \mathbf{z}_{i} = f_{\boldsymbol{\theta}_i}(\mathbf{x}_i) = \sigma(\mathbf{W}_i\mathbf{x}_i + \mathbf{b}_i),
\end{equation}
where $f_{\boldsymbol{\theta}_i}: \mathbb{R}^{d_i} \to \mathbb{R}^{d_{i+1}}$ with $\boldsymbol{\theta}_i=\{\mathbf{W}_i\in\mathbb{R}^{d_{i+1}\times d_i},\mathbf{b}_i\in\mathbb{R}^{d_{i+1}}\}$, and $\sigma$ is an activation. Unlike a standard FC layer that directly sets $\mathbf{x}_{i+1}=\mathbf{z}_i$, CAS further refines $\mathbf{x}_i$ before producing the final output.

\textbf{Adaptive Reweighting.} We predict feature-wise importance for $\mathbf{x}_i$ from $\mathbf{z}_i$:
\begin{equation}
    \mathbf{e}_i = g_{\boldsymbol{\phi}_i}(\mathbf{z}_{i}) = \texttt{Sigmoid}(\mathbf{W}'_i\mathbf{z}_{i} + \mathbf{b}'_i),
\end{equation}
where $g_{\boldsymbol{\phi}_i}: \mathbb{R}^{d_{i+1}} \to \mathbb{R}^{d_i}$ with $\boldsymbol{\phi}_i=\{\mathbf{W}'_i\in\mathbb{R}^{d_i\times d_{i+1}},\mathbf{b}'_i\in\mathbb{R}^{d_i}\}$. Each element of $\mathbf{e}_i$ indicates the importance of the corresponding feature in $\mathbf{x}_i$; smaller values imply stronger denoising. We then reweight the input by
\begin{equation}
    \mathbf{\tilde{x}}_i = \mathbf{x}_i \odot 2\mathbf{e}_i,
\end{equation}
where scaling by $2$ keeps weights in $(0,2)$ with expectation $1$, avoiding overly small activations and mitigating gradient vanishing (also used in~\cite{lhuc}).

\textbf{Cycle-Forward.} Using $\mathbf{\tilde{x}}_i$, we recompute the layer output with shared parameters:
\begin{equation}
    \mathbf{z}'_{i} = f_{\boldsymbol{\theta}_i}(\mathbf{\tilde{x}}_i) = \sigma(\mathbf{W}_i\mathbf{\tilde{x}}_i + \mathbf{b}_i).
\end{equation}
$\mathbf{z}'_i$ can be fed back to start another cycle, but we use a single cycle to balance effectiveness and efficiency, and set $\mathbf{x}_{i+1}=\mathbf{z}'_i$.

By inserting CAS layers in each user/item tower, we achieve per-layer self-correction and denoising without introducing new features, improving the robustness of the learned representations.

\begin{figure*}[t]
  \centering
  \vspace{-1em}
  \includegraphics[width=0.9\linewidth]{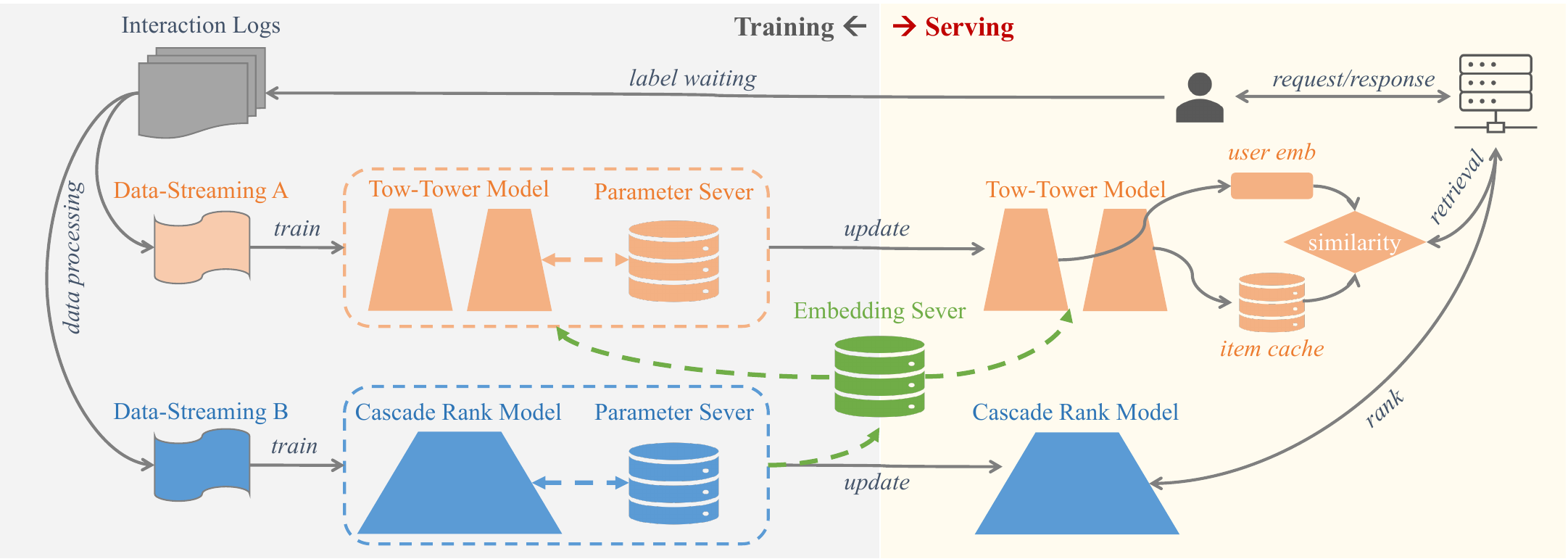} 
  \vspace{-1em}
  \caption{An overview of the online learning framework in our system. 
  } 
  \label{fig:online} 
  \vspace{-1em}
\end{figure*}

\vspace{-0.5em}
\subsection{Cross-Tower Synchronization}

In a two-tower model, the user and item towers are independent before similarity computation, which limits capacity and makes embedding alignment harder~\cite{WangYM000M22DirectAU}. Prior work enables cross-tower information exchange via implicit interactions~\cite{dat,IntTower}, but does not explicitly leverage representations from the partner tower. In online learning, where users/items and interactions continually evolve, implicit alignment can be insufficient. We therefore propose \textit{Cross-Tower Synchronization} (CTS), an explicit and lightweight exchange mechanism (orange part in Figure~\ref{fig:CS3}).

CTS maintains cross vectors $\mathbf{c}^t_u$ and $\mathbf{c}^t_v$ that cache positive representations from the partner tower for each user and item up to timestep $t>0$, initialized as $\mathbf{c}^0_u=\mathbf{0}$ and $\mathbf{c}^0_v=\mathbf{0}$. Given an interaction $y$ between user $u$ and item $v$ at timestep $t$, the tower outputs are
\begin{equation}
\label{eq:tower}
\begin{aligned}
    \mathbf{u}^t &= U_{\boldsymbol{\theta}_u}(\mathbf{x}^t_u, \mathbf{c}^{t-1}_u,\mathbf{s}^{t-1}_u), \\
    \mathbf{v}^t &= V_{\boldsymbol{\theta}_v}(\mathbf{x}^t_v, \mathbf{c}^{t-1}_v,\mathbf{s}^{t-1}_v),
\end{aligned}
\end{equation}
where $\mathbf{x}^t_u,\mathbf{x}^t_v$ are the original tower features, and $\mathbf{s}^{t-1}_u,\mathbf{s}^{t-1}_v$ come from CMS (Section~\ref{sec:CMS}). Thus, $\mathbf{u}^t$ and $\mathbf{v}^t$ are conditioned on cached partner information from the previous step.

After computing $\mathbf{u}^t$ and $\mathbf{v}^t$, CTS updates the cross vectors using exponential moving average (EMA), which is widely used for gradient-free updates~\cite{vqvae,RQVAE}:
\begin{equation}
\begin{aligned}
    \mathbf{c}_u^{t} &= \begin{cases}
            \alpha \mathbf{c}_u^{t-1} + (1-\alpha)\mathbf{v}^{t}, & \text{if } y=1 \\
            \mathbf{c}_u^{t-1}, & \text{otherwise}
        \end{cases}\\
    \mathbf{c}_v^{t} &= \begin{cases}
            \alpha \mathbf{c}_v^{t-1} + (1-\alpha)\mathbf{u}^{t}, & \text{if } y=1 \\
            \mathbf{c}_v^{t-1}, & \text{otherwise}
        \end{cases}
\end{aligned}
\end{equation}
where $\alpha\in[0,1]$ controls the smoothing. We only use representations from positive interactions ($y=1$) to update $\mathbf{c}^t_u$ and $\mathbf{c}^t_v$, capturing personalized user interests and item characteristics.

By injecting partner-tower representations into each tower's inputs, CTS enables explicit cross-tower interaction and improves user--item space alignment. Moreover, in online learning, CTS updates these cached signals in real time, facilitating faster adaptation to distribution shifts in the partner tower outputs.



\vspace{-0.5em}
\subsection{Cascade-Model Sharing}
\label{sec:CMS}

Two-tower retrievers trade capacity for efficiency, leaving a clear representation gap to downstream cascade rankers that exploit user--item cross features~\cite{WangFFW17DeepCross,BianWRPZXSZCMLX22CAN,YuYWZMLJK22CFFNN} or target-aware sequential modeling~\cite{SIM,TWIN,SDIM,ETA}, which can hurt end-to-end performance in multi-stage pipelines. Knowledge distillation transfers signals from strong teachers to two-tower students~\cite{TangW18, reddi2021rankdistil}, but it does not address the architectural limitation and often under-utilizes cascade models. \textit{Cascade-Model Sharing} (CMS) instead reuses intermediate outputs of cascade rankers as additional inputs to the two-tower model (blue part in Figure~\ref{fig:CS3}), enabling cross-stage computation sharing and reuse to improve two-tower capacity and pipeline consistency.

Analogous to CTS, CMS maintains cascade vectors $\mathbf{s}^{t}_u$ and $\mathbf{s}^{t}_v$ that cache cascade-model outputs for each user and item, initialized as $\mathbf{s}^0_u=\mathbf{0}$ and $\mathbf{s}^0_v=\mathbf{0}$. At timestep $t$, $\mathbf{s}^{t-1}_u$ and $\mathbf{s}^{t-1}_v$ are fed into the user/item towers together with $\mathbf{x}^t_u,\mathbf{x}^t_v$ and $\mathbf{c}^{t-1}_u,\mathbf{c}^{t-1}_v$ as in Equation~\ref{eq:tower}.

Let $\mathbf{h}^{t}_{uv}$ denote an intermediate representation produced by a cascade ranking model for pair $(u,v)$ at timestep $t$ (e.g., the penultimate FC output before the final prediction layer). CMS updates the cached vectors by EMA:
\begin{equation}
\begin{aligned}
    \mathbf{s}_u^{t} &= \beta \mathbf{s}_u^{t-1} + (1-\beta)\mathbf{h}^{t}_{uv},\\
    \mathbf{s}_v^{t} &= \beta \mathbf{s}_v^{t-1} + (1-\beta)\mathbf{h}^{t}_{uv},
\end{aligned}
\end{equation}
where $\beta\in[0,1]$ controls smoothing. Unlike CTS (updated only by positive interactions), CMS treats both positive and negative $\mathbf{h}^{t}_{uv}$ as useful, since cascade models encode richer knowledge.

CMS improves two-tower representations via cross-stage sharing while keeping the pipeline consistent. The cascade model can be jointly trained with the retriever or trained separately on different sample streams.

\begin{table*}[htbp]
\vspace{-1em}
\caption{Overall Performance of Offline Experiments}
\label{tab:offline}
\vspace{-1em}
\centering
\begin{tabular}{lcccccc}
\toprule
\multirow{2}{*}{Model} & \multicolumn{2}{c}{TaobaoAd}                 & \multicolumn{2}{c}{KuaiRand}                  & \multicolumn{2}{c}{RecSys2017}   \\
\cmidrule(lr){2-3}\cmidrule(lr){4-5}\cmidrule(lr){6-7}
                       & {AUC $\uparrow$}   & {LogLoss $\downarrow$}  & {AUC $\uparrow$}   & {LogLoss $\downarrow$}  & {AUC $\uparrow$}   & {LogLoss $\downarrow$} \\

\midrule
\rowcolor{gray!10}
DSSM\cite{dssm2013}    & 0.6194$\pm$.0028             & 0.2289$\pm$.0002                  & 0.6646$\pm$.0027             & 0.6763$\pm$.0003             & 0.6855$\pm$.0073             & 0.6707$\pm$.0039 \\
\hspace{1em}+ CAS      & 0.6378$\pm$.0015             & 0.2271$\pm$.0002                  & \underline{0.7416$\pm$.0004} & \underline{0.5759$\pm$.0002} & 0.7093$\pm$.0245             & 0.6864$\pm$.0129 \\
\hspace{1em}+ CTS      & 0.6506$\pm$.0018             & 0.2258$\pm$.0001                  & 0.7095$\pm$.0006             & 0.6122$\pm$.0004             & \underline{0.7752$\pm$.0182} & \underline{0.6166$\pm$.0051} \\
\hspace{1em}+ CMS      & \underline{0.6632$\pm$.0011} & \underline{0.2253$\pm$.0001}      & 0.7100$\pm$.0016             & 0.6094$\pm$.0002             & 0.7417$\pm$.0130             & 0.6353$\pm$.0032 \\
\rowcolor{gray!30}
\hspace{1em}+ CS3      & \textbf{0.6855$\pm$.0005*}   & \textbf{0.2198$\pm$.0001*}        & \textbf{0.7484$\pm$.0008*}   & \textbf{0.5731$\pm$.0004*}   & \textbf{0.8380$\pm$.0038*}   & \textbf{0.5308$\pm$.0015*} \\
\midrule
\rowcolor{gray!10}
IntTower\cite{IntTower}& 0.6507$\pm$.0004             & 0.2255$\pm$.0001                  & 0.7503$\pm$.0016             & 0.5782$\pm$.0057             & 0.8178$\pm$.0057             & 0.6429$\pm$.0203 \\
\hspace{1em}+ CAS      & 0.6541$\pm$.0005             & 0.2251$\pm$.0002                  & \underline{0.7580$\pm$.0002} & \underline{0.5636$\pm$.0002} & \underline{0.8511$\pm$.0018} & \underline{0.4980$\pm$.0015} \\
\hspace{1em}+ CTS      & \underline{0.6825$\pm$.0008} & \underline{0.2213$\pm$.0002}      & 0.7527$\pm$.0002             & 0.5686$\pm$.0005             & 0.8445$\pm$.0036             & 0.5975$\pm$.0237 \\
\hspace{1em}+ CMS      & 0.6745$\pm$.0014             & 0.2217$\pm$.0005                  & 0.7571$\pm$.0003             & 0.5787$\pm$.0006             & 0.8236$\pm$.0062             & 0.6244$\pm$.0149 \\
\rowcolor{gray!30}
\hspace{1em}+ CS3      & \textbf{0.6895$\pm$.0003*}   & \textbf{0.2186$\pm$.0001*}        & \textbf{0.7615$\pm$.0002*}   & \textbf{0.5572$\pm$.0002*}   & \textbf{0.8657$\pm$.0030*}   & \textbf{0.4888$\pm$.0090*} \\
\midrule
\rowcolor{gray!10}
IHM-DAT\cite{ihm}      & 0.6302$\pm$.0037             & 0.2278$\pm$.0002                  & 0.7059$\pm$.0097             & 0.6387$\pm$.0011             & 0.7694$\pm$.0031             & 0.6096$\pm$.0019 \\
\hspace{1em}+ CAS      & \underline{0.6544$\pm$.0013} & \underline{0.2247$\pm$.0001}      & \underline{0.7513$\pm$.0024} & 0.6040$\pm$.0012             & \underline{0.8529$\pm$.0011} & \underline{0.5413$\pm$.0017} \\
\hspace{1em}+ CTS      & 0.6478$\pm$.0017             & 0.2255$\pm$.0001                  & 0.7083$\pm$.0003             & 0.6037$\pm$.0006             & 0.7848$\pm$.0036             & 0.6044$\pm$.0035 \\
\hspace{1em}+ CMS      & 0.6532$\pm$.0008             & 0.2253$\pm$.0002                  & 0.7103$\pm$.0044             & \underline{0.6022$\pm$.0019} & 0.7925$\pm$.0021             & 0.6007$\pm$.0017 \\
\rowcolor{gray!30}
\hspace{1em}+ CS3      & \textbf{0.6783$\pm$.0021*}   & \textbf{0.2216$\pm$.0002*}        & \textbf{0.7556$\pm$.0035*}   & \textbf{0.5723$\pm$.0019*}   & \textbf{0.8660$\pm$.0014*}   & \textbf{0.5307$\pm$.0026*} \\
\midrule
\rowcolor{gray!10}
RCG\cite{RCG}          & 0.6680$\pm$.0001             & 0.2217$\pm$.0001                  & 0.7814$\pm$.0042             & 0.5569$\pm$.0028             & 0.7870$\pm$.0026             & 0.5697$\pm$.0031 \\
\hspace{1em}+ CAS      & 0.6741$\pm$.0008             & 0.2213$\pm$.0001                  & \underline{0.8195$\pm$.0063} & \underline{0.5361$\pm$.0029} & \underline{0.8390$\pm$.0027} & \underline{0.5348$\pm$.0042} \\
\hspace{1em}+ CTS      & \underline{0.6764$\pm$.0002} & \underline{0.2210$\pm$.0001}      & 0.7931$\pm$.0025             & 0.5475$\pm$.0023             & 0.7954$\pm$.0049             & 0.5854$\pm$.0041 \\
\hspace{1em}+ CMS      & 0.6725$\pm$.0009             & 0.2212$\pm$.0001                  & 0.7906$\pm$.0013             & 0.5561$\pm$.0010             & 0.8061$\pm$.0121             & 0.5689$\pm$.0065 \\
\rowcolor{gray!30}
\hspace{1em}+ CS3      & \textbf{0.6860$\pm$.0012*}   & \textbf{0.2203$\pm$.0003*}        & \textbf{0.8304$\pm$.0009*}   & \textbf{0.5241$\pm$.0012*}   & \textbf{0.8676$\pm$.0031*}   & \textbf{0.5132$\pm$.0073*} \\
\bottomrule
\end{tabular}
\begin{minipage}{50em} \footnotesize
 Results for the base and CS3-enhanced two-tower models are highlighted in gray. In each comparison group, boldface indicates the \textbf{best result}, underline indicates the \underline{second-best}, and an asterisk (*) denotes statistically significant improvement over the base model with $p<0.05$.
\end{minipage}
\vspace{-1em}
\end{table*}

\vspace{-0.5em}
\subsection{Implementation in Online Learning}
\label{sec:implementation}

We present an efficient CS3 implementation for online learning in our production system. The end-to-end training and serving workflow is shown in Figure~\ref{fig:online}.

\subsubsection{Online Learning}

In online learning, displayed items undergo a label waiting period; once labels are ready, logs are ingested in real time for training, and updated parameters are periodically synchronized to online servers. This loop typically finishes within an hour (often $\sim$30 minutes), and is widely used in industry to improve freshness and user experience~\cite{DBLP:conf/recsys/0004ZZWZTZZWWC22,cao2024momentcrossnextgenerationrealtimecrossdomain}. Compared to offline training, online learning imposes stricter latency constraints and complicates optimization, especially for CS3 which requires efficient information exchange across towers and stages.

\subsubsection{Parameter Server}

To support large-scale real-time training, we use a distributed \textit{Parameter Server} (ParSvr) to update and synchronize model parameters, including sparse embeddings (e.g., per-\texttt{user\_id}). Although these embeddings resemble CTS cross vectors in form, they are optimized by gradients, whereas cross vectors are updated by EMA. We therefore cache CTS cross vectors in ParSvr and implement EMA updates via custom gradients. All parameters are synchronized across nodes and periodically exported for online deployment.

\subsubsection{Embedding Server}

ParSvr is scoped to a single training job and cannot share cached vectors across different training processes, which prevents CMS from caching cascade vectors unless the two-tower and cascade models are jointly trained. In practice, these models often use different data streams and sampling strategies. We thus adopt an independent \textit{Embedding Server} (EmbSvr) to store cascade vectors. EmbSvr is a distributed key--value service specialized for embedding storage and higher QPS under the same hardware. Cascade-model outputs are cached in EmbSvr with \texttt{user\_id}/\texttt{item\_id} as keys and updated by EMA; the two-tower model retrieves them during both training and serving.

\subsubsection{Efficiency}

CTS and CMS add only two extra input vectors, incurring negligible compute overhead. CTS cross vectors are updated and synchronized with other sparse features (in ParSvr during training and stored locally during serving), with no observable latency impact. CMS cascade vectors are fetched from EmbSvr in both training and serving; EmbSvr maintains p99 latency below 5\,ms, and its accesses are parallelized with other feature processing, yielding negligible end-to-end overhead.

CAS increases per-tower computation, but retrieval latency is dominated by large-scale similarity search, which is unchanged. Moreover, item embeddings are precomputed and cached, so the main overhead comes from the real-time user tower. In our deployment, applying CAS to all FC layers except the input layer reduces the retriever service QPS by less than 1\%.

\vspace{-0.5em}
\section{Experiments}

\subsection{Experimental Settings}

\subsubsection{Offline}
We conducted offline experiments on three public datasets from diverse industrial domains with both user and item features: TaobaoAd\footnote{\url{https://tianchi.aliyun.com/dataset/56?lang=en-us}}, KuaiRand\footnote{\url{https://kuairand.com/}}, and RecSys2017\footnote{\url{https://www.recsyschallenge.com/2017/}}. 
CS3 was applied to several two-tower architectures to assess its effectiveness and generalizability, including DSSM~\cite{dssm2013}, IntTower~\cite{IntTower}, IHM-DAT~\cite{ihm}, and transformer-based RCG~\cite{RCG}. 
To generate the CMS cascade vectors, we employed a feedforward neural network with four hidden layers for user and item feature modeling, which was jointly trained with the two-tower models. Codes is available at \url{https://github.com/lixiangwang/CS3Rec}.

\subsubsection{Online}
We conduct A/B tests across three different business scenarios in our online system, which serves hundreds of millions of users. In each scenario, we enhance the two-tower retrieval model providing the largest pool of candidates for the subsequent stages to evaluate the real-world performance of CS3.

\vspace{-0.5em}
\subsection{Offline Experiments}

Table~\ref{tab:offline} reports mean AUC and LogLoss (with standard errors) on test sets over five random seeds. We summarize the key findings:

1) Across datasets and two-tower backbones, adding CAS, CTS, or CMS consistently improves performance, validating each CS3 module.  
2) On KuaiRand, CAS brings the largest gains, because its self-correction/denoising better exploits the dataset's rich user/item features.  
3) DSSM+CS3 outperforms IntTower and IHM-DAT, showing CS3 surpasses strong two-tower optimization baselines.  
4) Adding CTS further improves IHM-DAT, suggesting explicit cross-tower exchange is more effective than its implicit alignment.  
5) Adding CMS improves IntTower, indicating that downstream-stage signals provide additional benefits beyond interaction modeling within the two-tower structure.  
6) CS3 consistently boosts diverse architectures, including recent transformer-based retrievers, demonstrating strong generalization.

\begin{table}[htbp]
\caption{Online performance of CS3 on Scenario A}
\label{tab:onlineA}
\vspace{-1em}
\centering
\setlength{\tabcolsep}{10pt}
\begin{tabular}{lrr}
\toprule
 {Method} & {Revenue} & {DAC} \\
\midrule
BASE                          & 0.000\%  & 0.000\%  \\
+ CAS                         & +1.677\%  & +0.144\%  \\
+ CAS\&CMS                   & +7.880\%  & +0.435\%  \\
+ CAS\&CMS\&CTS(CS3)         & +8.356\%  & +0.468\%  \\
\bottomrule
\end{tabular}
\end{table}

\begin{table}[htbp]
\caption{Online improvement of CS3 across scenarios}
\label{tab:online3}
\vspace{-1em}
\centering
\setlength{\tabcolsep}{8pt}
\begin{tabular}{lccc}
\toprule
BASE+CS3 & {ScenarioA} & {ScenarioB} & {ScenarioC} \\
\midrule
Revenue      & +8.356\% & +1.366\% & +2.177\% \\
DAC      & +0.468\% & +0.143\% & +0.228\% \\
QPS      & -0.589\% & -0.388\% & -0.456\% \\
\bottomrule
\end{tabular}
\vspace{-1em}
\end{table}

\vspace{-0.5em}
\subsection{Online A/B Tests}

We run online A/B tests in a large-scale advertising recommender serving over 400 million daily active users. CS3 is first deployed in Scenario A and achieves significant gains (Table~\ref{tab:onlineA}). To evaluate generalizability, we further deploy CS3 in two additional business scenarios, also obtaining strong improvements (Table~\ref{tab:online3}). We report advertising revenue, Daily Active Customers (DAC), and the QPS reduction after applying CS3. The main observations are:

1) CS3 consistently improves online performance; among the components, CMS contributes the largest gains by improving consistency between the two-tower retriever and downstream rankers, better leveraging increased capacity.  
2) Improvements in ad recommendation translate into higher revenue and better user experience/conversion, which is especially valuable at this scale.  
3) CS3 yields consistent gains across all scenarios, demonstrating strong generalizability.  
4) CS3 slightly reduces QPS, mainly due to extra user-tower computation from CAS (Section~\ref{sec:implementation}); the latency impact is minimal and outweighed by the effectiveness gains.

All tests are run on the two-tower retriever (the largest candidate source in each scenario) using 10\% real traffic for more than 7 days. The improved model is then rolled out to 100\% traffic to replace BASE. Each scenario has daily peak QPS above 200k. We continue to refine CS3 and extend it to more models and scenarios.




\vspace{-0.5em}
\section{Conclusion}

We propose CS3, a general framework for strengthening two-tower recommenders. CS3 consists of three components---CAS, CTS, and CMS---that let each tower leverage signals from itself, its partner tower, and a stronger cascade model, improving representation quality and cross-tower/stage synergy. We also present an efficient online-learning implementation and deploy CS3 in a large-scale advertising system. Extensive offline experiments and online A/B tests demonstrate consistent effectiveness and strong generalizability.


\bibliographystyle{ACM-Reference-Format}
\balance
\bibliography{reference}

\end{document}